\documentclass[pre,preprint,aps,showpacs,unsortedaddress]{revtex4}
\usepackage{epsfig}
\usepackage{graphics}
\begin{document}

\title{Maximum likelihood approach for several stochastic volatility models}

\author{Jordi Camprodon}
\email{jordicampro@gmail.com}
\author{Josep Perell\'o}
\email{josep.perello@ub.edu}
\affiliation{Departament de F\'{\i}sica Fonamental, Universitat de Barcelona,\\Diagonal, 647, E-08028 Barcelona, Spain}
\date{\today}
\begin{abstract}
Volatility measures the amplitude of price fluctuations. Despite it is one of the most important quantities in finance, volatility is not directly observable. Here we apply a maximum likelihood method which assumes that price and volatility follow a two-dimensional diffusion process where volatility is the stochastic diffusion coefficient of the log-price dynamics. We apply this method to the simplest versions of the expOU, the OU and the Heston stochastic volatility models and we study their performance in terms of the log-price probability, the volatility probability, and its Mean First-Passage Time. The approach has some predictive power on the future returns amplitude by only knowing current volatility. The assumed models do not consider long-range volatility auto-correlation and the asymmetric return-volatility cross-correlation but the method still arises very naturally these two important stylized facts. We apply the method to different market indexes and with a good performance in all cases.
\end{abstract}
\maketitle

\section{Introduction}
 \label{sec_intro}

Volatility is a magnitude aiming to capture how big is the amplitude of price return fluctuations~\cite{hull,bouchaud2003theory}. It is associated with the risk of holding an asset stating that the higher the volatility the riskier the market price. Investors pay sometimes more attention to volatility than to the price level or the current trend of a stock. The role of volatility becomes even more crucial when trading with financial derivatives like options since the value of volatility almost fully determines the price of this sort of contracts~\cite{hull,bouchaud2003theory}. However, the volatility itself is not directly observed and the financial markets and their actors lack of an unique consensus for providing its value. 

Therefore, there is no other choice than trying to infer in some way or another the value of volatility from price time series. In practice, this means that it is necessary to first assume a model governing financial asset dynamics and second to extract volatility value from data time series under the perspective of the model dynamics considered. 

The physicist Osborne proposed the Geometric Brownian Motion model (GBM) in 1959~\cite{osborne1959brownian}. The GBM difussion process drives the logarithmic price changes with a constant diffusion coefficient typically called volatility. In this case, computing market volatility first means to calculate the standard deviation of the logarithmic price changes over time periods of length $\Delta t$. And, secondly, volatility would then be the ratio between the standard deviation and the square root of $\Delta t$ since we are implicitly assuming the GBM difussion model. 

Further studies in financial data have led to establish that the GBM is very incomplete~\cite{bouchaud2003theory} and it appears to be unable to explain quite a long list of stylized facts observed in financial markets~\cite{bouchaud2003theory,cont2001empirical}. Specially during the last two decades, several models have been proposed with the aim of capturing (i) the existence of fatter tails in the log-price fluctuations, and (ii) the presence of non-trivial memory in the market dynamics~\cite{bouchaud2003theory}. A very natural improvement of the GBM is to consider volatility as a random process following another continuous time diffusion process~\cite{hullpaper,scott,wiggins,Barndorff,saichev,bacry,calvet,delpini}. The price and the hidden Markov process for the volatility therefore configure a two-dimensional difussion process and the approach belongs to the so-called stochastic volatility (SV) modeling~\cite{Ghysels,fouque2000derivatives}. The approach is analogous to random diffusion modeling which describes dynamics of particles in random media and applicable to a large variety of phenomena in statistical physics and condensed matter~\cite{ben-avraham}.

Among the existing SV models~\cite{Ghysels,fouque2000derivatives,perello2004comparison,delpini}, the most basic ones are the Ornstein-Uhlenbeck (OU)~\cite{stein,perelloijtaf,bormettiijtaf}, the Heston model~\cite{heston1993closed,dragulescu,bormetti2010} being in fact a Feller process, and the exponential Ornstein-Uhlenbeck (expOU)~\cite{masoliver2006multiple,sircar,bormetti2008}. With the aim of extracting volatility from financial markets data, the current work develops much further the maximum likelihood (ML) estimation applied to the expOU model in Ref.~\cite{eisler2007volatility} by one of us. We here extend the methodology to the OU and Heston SV models but we also study some of the most important statistical features observed in financial markets~\cite{bouchaud2003theory,cont2001empirical,delpini}: the return and volatility probability densities (pdf's), the volatility auto-correlation and the leverage correlation, and the Mean First-Passage Time. For doing all these, we use eight daily indexes: Dow Jones Industrial Average (DJI), Standard and Poor's-500 (S\&P), German index DAX, Japanese index NIKKEI, American index NASDAQ, British index FTSE-100, Spanish index IBEX-35 and French index CAC-40. We also provide the method abilities of predicting future absolute value of price returns knowing today's volatility.

This paper is divided into five sections. In Section \ref{sec_stochvol} we present the SV models and their main characteristics, while in Section \ref{sec_maxlike} we show the ML approach. In Section \ref{sec_simres} we provide results obtained from our algorithm. Conclusions are left to Section~\ref{sec_conclusions}.

\section{The stochastic volatility market models and basic volatility estimators}
 \label{sec_stochvol}

The starting point of any SV model is the GBM model~\cite{osborne1959brownian}
\begin{equation}
 \label{eq_1}
\frac{dS(t)}{S(t)}=\mu dt + \sigma dW_1 (t),
\end{equation}
where $dW_1(t)$ corresponds to a Wiener noise (i.e., a zero mean and unit variance Gaussian process), $S(t)$ is a financial price or the value of an index, $\mu$ is the drift and $\sigma$ is the volatility. 

If we define the zero-mean return $X(t)$ as
\begin{equation}
 \label{eq_2}
X(t)=\ln\left(\frac{S(t+t_0)}{S(t_0)}\right)- \left\langle \ln\left(\frac{S(t+t_0)}{S(t_0)}\right) \right\rangle,
\end{equation}
where $t_0$ is the initial time. Let us note that $X(t)$ assumes independent and stationary increments in the financial time series since Osborne's work in 1959~\cite{osborne1959brownian}. We can however rewrite Eq.~(\ref{eq_1}) as follows
\begin{equation}
 \label{eq_3}
dX(t) = \sigma(t) dW_1 (t).
\end{equation}
The term $\sigma$ was initially considered to be constant. However, most of the existing market models nowadays assumes that the term $\sigma$ --also called volatility-- is a time varying variable.

SV models assume that the volatility is a hidden Markov process $\sigma(t)=f(Y(t))$ where $Y(t)$ obeys a subordinated diffusive stochastic differential equation. Under this perspective, the two-dimensional dynamics reads~\cite{fouque2000derivatives}
\vspace{-0.2cm}
\begin{eqnarray}
&&dX(t) = f(Y(t)) dW_1 (t),\label{eq_4}\\
&&dY(t) = -g(Y(t))dt +h(Y(t)) dW_2 (t),
\label{eq_5}
\end{eqnarray}
where $W_i(t)$ $(i=1,2)$ are Wiener processes that may or not be independent. As $f(y)$ is always defined as a monotonically increasing function, $Y(t)$ is sometimes also called volatility. As shown in Tab. \ref{table_fgh}, each model has its own expressions of $f(y)$, $g(y)$ and $h(y)$. The proposed models in the literature change in terms of these functions but in general there is a wide consensus to consider process with a (negative) mean reverting force that leads the probability density function of the volatility to a stationary solution when time is sufficiently large.

Let us focus on the volatility estimation procedures. As a first approximation and as mentioned in the introduction, the volatility can be viewed as the standard deviation of the empirical daily zero-mean return changes
$$
\sigma_{GBM}=\sqrt{\frac{\langle \Delta X(t)^2\rangle}{\Delta t}}.
$$
As we are considering daily data, we are assuming discrete time increments $\Delta t=1\,\mbox{day}$ and discrete return increments $\Delta X(t)= X(t+ 1\,{\rm day})-X(t)$. In such a case, we are implicitly assuming the GBM provided by Eq.~(\ref{eq_3}) with constant volatility in daily units.

As a second level of approximation we allow for time varying volatility. Observing Eq.~(\ref{eq_3}), we now define volatility as 
\begin{equation}
 \label{eq_10}
\sigma_{\mbox{\small prop}} (t) = \frac{|\Delta X(t)|}{\left\langle |\Delta W_1(t)| \right\rangle},
\end{equation}
and we have different volatility for different days. However, the volatility obtained has a skewed stationary probability density inconsistent with volatility modeling as discussed in Refs.~\cite{bouchaud2003theory,masoliver2006multiple}. 

A third possibility is to compute a deconvoluted volatility~\cite{masoliver2006multiple}
\begin{equation}
 \label{eq_11}
\sigma_{\mbox{\small decon}}(t) = \left| \frac{\Delta X(t)}{\Delta W_1(t)} \right|,
\end{equation}
which does not show a skewed probability density for the volatility but its greatest drawback is that estimated volatility appears to be a very noisy signal (see for instance Refs.~\cite{bouchaud2003theory,masoliver2006multiple,silva,stanley} for alternative approaches and further discussions).

\begin{table}[t]
\caption{Volatility expressions in terms of $f(y)$, $g(y)$ and $h(y)$ appearing in Eq.~(\ref{eq_5}). These models have three constants: the normal level of volatility $m$, the driving force $\alpha$ that drives volatility to $m$, and the amplitude of volatility fluctuations $k$ often called volatility-of-volatility~\cite{fouque2000derivatives}.}
\vspace{0.3cm}
\label{table_fgh}
\begin{center}
\begin{tabular}{llll}
\hline\hline\noalign{\smallskip}
 & expOU & OU & Heston \\
\noalign{\smallskip}\hline\noalign{\smallskip}
$f(y)$ & $m e^y$ & $y$ & $y^{1/2}$ \\
$g(y)$ & $\alpha y$ & $\alpha(y-m)$ & $\alpha(y-m)$ \\
$h(y)$ & $k$ & $k$ & $ky^{1/2}$ \\
\noalign{\smallskip}\hline
\hline
\end{tabular}
\end{center}
\end{table}

\section{Maximum likelihood approach}
\label{sec_maxlike}
 
We here briefly present the methodology proposed in Ref.~\cite{eisler2007volatility} that allows us to have some criteria for choosing the best values of the random realization $\Delta W_1$. Naively speaking, the method represents an improvement of the deconvoluted volatility $\sigma_{\mbox{\small decon}}$ estimator using a ML methodology. 

To explain the procedure it is more convenient to work with the discrete time version of the model. To this end, suppose that $\Delta t$ is a small time step and that the driving noises in Eqs. (\ref{eq_4})-(\ref{eq_5}) can be approximated by
\begin{equation}
dW_{i}(t) \approx\varepsilon_{i}(t)\sqrt{\Delta t}, \qquad (i=1,2),
\label{a3}
\end{equation}
where $\varepsilon_i(t)$ are independent standard Gaussian processes with zero mean and unit variance. The discrete time equations of the model describing increments of $X(t)$ and $Y(t)$ thus read
\begin{eqnarray}
\Delta X(t) &=& f(Y(t))\varepsilon_{1}(t)\sqrt{\Delta t}\, \label{a4} \\
\Delta Y(t) &=& -g(Y(t))\Delta t+h(Y(t))\varepsilon_{2}(t)\sqrt{\Delta t}\;
\label{a5}
\end{eqnarray}
where $\Delta X(t)= X(t+ \Delta t)-X(t)$ and $\Delta Y(t)= Y(t+ \Delta t)-Y(t)$. From Eqs.~(\ref{a4})--(\ref{a5}), we can get
\begin{eqnarray}
\label{a6}
\varepsilon_{1}(t) &=& \frac{\Delta X(t)}{f(Y(t))\sqrt{\Delta t} },\\
\label{a7}
\varepsilon_{2}(t)&=&\frac{\Delta Y(t)+g(Y(t))\Delta t}{h(Y(t))\sqrt{\Delta t}}.
\end{eqnarray}
For simplicity we assume that $\varepsilon_1$ and $\varepsilon_2$ are independent standard Gaussians. We will discuss in Section \ref{sec_corr} that our methodology does not need to consider negative cross-correlation among these two Gaussian variables as discussed for instance by the models studied in Refs.\cite{delpini,leverage,bormetti2008}. Hence,
$$
{\rm P}(\varepsilon_1,\varepsilon_2)=(1/2\pi)
\exp\left[(\varepsilon_{1}^2+\varepsilon_{2}^2)/2\right].
$$
and this finally can be transformed into the conditional probability density function (pdf)
\begin{eqnarray}
{\rm P}(X(\tau),Y(\tau)\vert X(\tau-\Delta t),Y(\tau-\Delta t))= \nonumber\\\frac{1/(2\pi\Delta t)}
{f(Y(\tau-\Delta t))h(Y(\tau-\Delta t))}
\exp\left[-\frac{\varepsilon_{1}^2(\tau-\Delta t)+\varepsilon_{2}^2(\tau-\Delta t)}{2}\right].
\label{a10}
\end{eqnarray}
by including the Jacobian of the transformation $(X(\tau),Y(\tau))\longrightarrow(\varepsilon_1(\tau-\Delta t),\varepsilon_2(\tau-\Delta t))$ defined by Eqs.~(\ref{a4})-(\ref{a5}).

For a given number of realizations, the probability of the set $\{\mathbf X,\mathbf Y\}$ for the period $(\tau=t,t-\Delta t,\dots,t-s)$ can be easily obtained. The Markov property of the process ensures that one can decompose the joint pdf of this set as a chain of products between conditional pdf's
\begin{eqnarray}
{\rm P}(\{\mathbf X,\mathbf Y\})=P(X(t-s),Y(t-s))\nonumber\\
\times \prod_{\tau=t+\Delta t-s}^{t} P(X(\tau),Y(\tau)|X(\tau-\Delta t),Y(\tau-\Delta t)).
\label{markov}
\end{eqnarray}
Substituting Eqs. (\ref{a6})-(\ref{a7}) into Eq.~(\ref{a10}) and inserting them into Eq. (\ref{markov}), we apply the chain of products between conditional pdf's and we finally get the joint pdf
\begin{eqnarray}
\ln {\rm P}(\{\mathbf X, \mathbf Y\})=-\frac{s\ln(2\pi\Delta t)}{\Delta t}\nonumber \\
-\sum_{\tau=t+\Delta t-s}^{t}[\ln f(Y(\tau-\Delta t))+ \ln h(Y(\tau-\Delta t))] \nonumber \\
+\ln {\rm P}(X(t-s), Y(t-s))
-\frac{1}{2}\sum_{\tau=t+\Delta t-s}^{t}\left[\frac{X(\tau)-X(\tau-\Delta t)}{f(Y(\tau-\Delta t))\Delta t}\right]^2 \Delta t \nonumber \\
-\frac{1}{2}\sum_{\tau=t+\Delta t-s}^{t}\left[\frac{Y(\tau)-Y(\tau-\Delta t)}{h(Y(\tau-\Delta t))\Delta t}+
\frac{g(Y(\tau-\Delta t))}{h(Y(\tau-\Delta t))}\right]^2\Delta t.
\label{a11}
\end{eqnarray}

We remind that our aim is to find a proper realization of the volatility $Y$ given a return $X$ and this will be done by applying a ML procedure to variable $Y$. For this reason, we will be able to omit three terms in Eq.~(\ref{a11}). The first summand comes from the normalization constant of the Gaussian distribution~(\ref{a10}). It appears in every conditional probability density and this is the reason for the factor $s/\Delta t$, which is the number of time steps between $t-s$ and $t$. The resulting term does not depend on the realization, so that we can neglect it for a maximization with respect to the set of realizations ${\mathbf Y}$. The second summand is mostly the sum of the Jacobian transformations of each transition probability. Stochastic volatility models assume that these $f$ and $g$ are continuous and monotonically increasing functions or even constants. Because of this, we can also neglect this term in the maximization procedure. The term $\ln {\rm P}(X(t-s),Y(t-s))$ is fixed by the initial conditions of the process. We could here assume a known initial return $X$ -- which can be set to zero -- and take a random $Y(t-s)$ following its stationary distribution. Therefore we would have ${\rm P}(X(t-s),Y(t-s))=\delta(X(t-s)-X)\ P_\mathrm{st}(Y(t-s))$. Had we taken another initial condition, the technique would have given equivalent results (we have checked this by using several initial distributions). For this reason and in order to improve the convergence of the ML estimate we have neglected also this contribution.

We can therefore write 
\begin{eqnarray}
\ln {\rm P}({\mathbf X}, {\mathbf Y})\simeq
-\frac{1}{2}\sum_{\tau=t+\Delta t-s}^{t}\left[\frac{\Delta X(\tau-\Delta t)}{f(Y(\tau-\Delta t))\Delta t}\right]^2 \Delta t \nonumber\\ 
-\frac{1}{2}\sum_{\tau=t+\Delta t-s}^{t}\left[\frac{Y(\tau-\Delta t)}{h(Y(\tau-\Delta t)}\Delta t+
\frac{g(Y(\tau-\Delta t))}{h(Y(\tau-\Delta t))}\right]^2\Delta t+\cdots\nonumber \\
\label{eq_6}
\end{eqnarray}
and omit the other three terms for the reasons summarized above (cf. Eq.~(\ref{a11})). Further details can be found in Ref.~\cite{eisler2007volatility}. 

Let us finally briefly provide an interpretation for the two remaining terms in Eq.~(\ref{eq_6}). The first term of Eq.~(\ref{eq_6}) measures the return variations with respect to the volatility. We notice that the higher the fluctuations are, the lower the contribution to the probability is. The second term computes the fluctuations of the volatility with respect to the volatility of the volatility. Again, the bigger this term, the lower the contribution.

\begin{figure}[t]
\vspace{-7cm}
\includegraphics[scale=0.45]{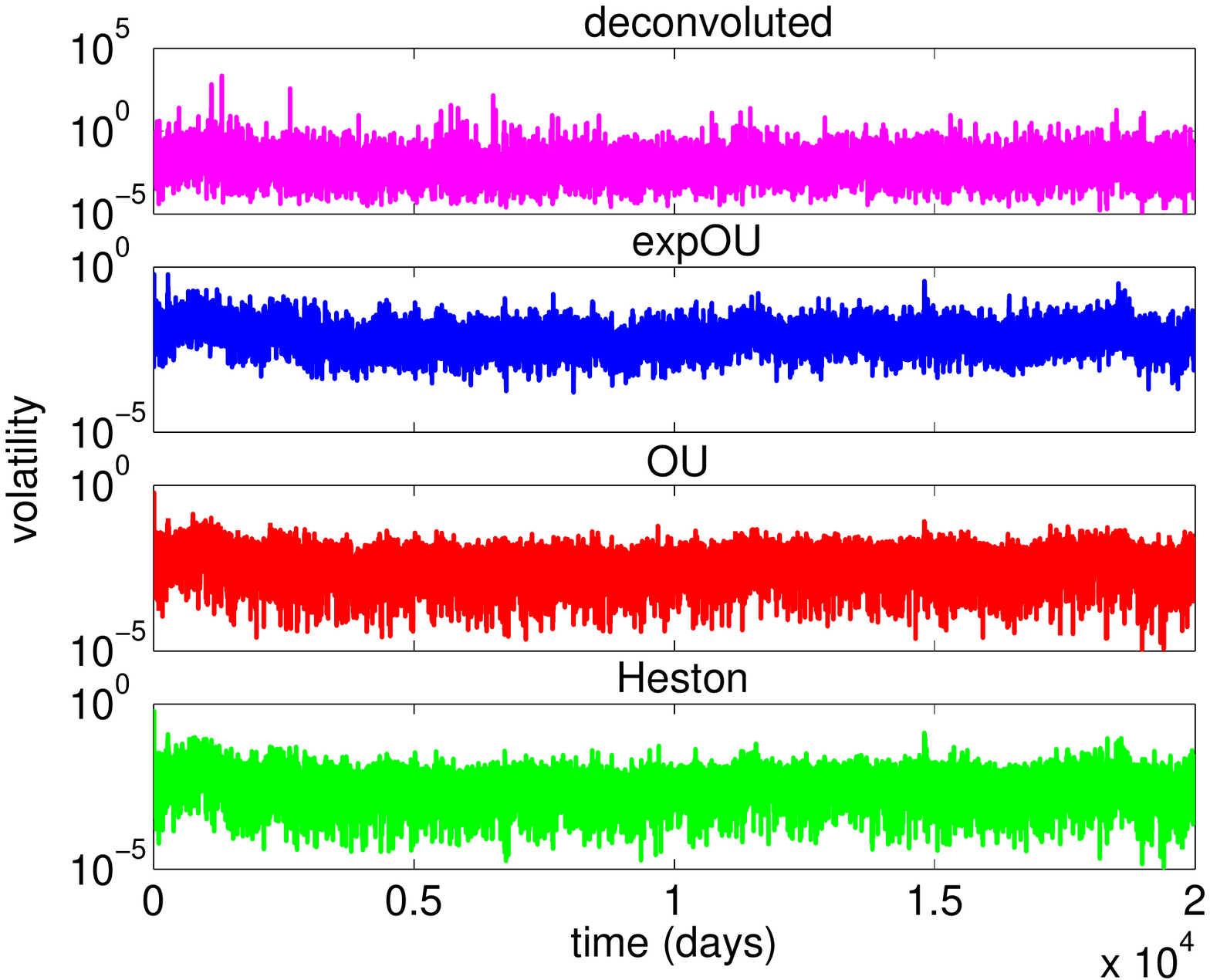}
\caption{A comparison between different volatilities as a function of time. On the top we observe the deconvoluted volatility $\sigma_{\mbox{\small decon}}$ computed using Eq. (\ref{eq_11}). The other plots show estimated volatilities $\sigma_{\mbox{\small est}}$ calculated for the three different models as explained in Section \ref{sect3a} and Eq.~(\ref{estimation}).}
\label{comparison_sigma} 
\end{figure}

\subsection{The Algorithm}
\label{sect3a}

As mentioned above, our goal is to find a proper realization of the volatility series $\mathbf Y$ given return series $\mathbf X$ which is directly observed and taken from empirical data. We then should however consider the following conditional probability of a single event
\begin{equation}
\label{eq_8}
\ln P(Y|X)=\ln P(X,Y)-\ln P(X).
\end{equation}
And as we solely want to maximize this probability for a fixed set of returns configuring a path, the second term can be neglected and therefore maximizing Eq.~(\ref{eq_8}) is equivalent to maximizing Eq.~(\ref{eq_6}). In practice, the method therefore computes different realizations of volatility variable for a given return path and ML estimation dictates that we should take the realization that makes bigger the probability given by Eq.~(\ref{eq_6}). The method filters the Wiener noise $\Delta W_1(t)$ and let us obtain an estimation $Y_{\mbox{\small est}}(t)$ of the hidden volatility for a given price return evolution.

Specifically, we have implemented an algorithm which sequentially follows the four steps:
\begin{enumerate}
\item Looking at Eq.~(\ref{eq_4}), we generate a simple realization of Y by taking
\begin{equation}
 \label{eq_9}
\bar{Y}_{\mbox{{\small est}}} (\tau) = f^{-1} \left( \left| \frac{\Delta X(\tau)}{\Delta W_1(\tau)} \right| \right)
\end{equation}
where $t-s \leq \tau \leq t$, with $\Delta X(\tau)=X(\tau+\Delta t)-X(\tau)$ taken from data, and $\Delta W_1(\tau)$ being a zero mean and unit variance Gaussian realization.

\item We substitute $\bar{Y}_{\mbox{\small est}}$ and $X$ into Eq.~(\ref{eq_6}) and we then compute the probability.

\item We iterate $I$ times the steps 1 and 2. We finally keep the realization that brings a higher probability in Eq.~(\ref{eq_6}) and define it as $Y_{\mbox{\small est}}(t)$.

\item Finally, the estimator of the volatility at time $t$ is 
\begin{equation}
\sigma_{\mbox{\small est}}(t) = f(Y_{\mbox{\small est}}(t)).
\label{estimation}
\end{equation}

\end{enumerate}
We observe that this procedure depends on $I$ and $s$. We have implemented the algorithm with $s=10$ and $I=100,000$. We have used these values because larger time window $s$ and a larger number $I$ of iterations do not improve the quality of our estimation. 

We observe that $\sigma_{\mbox{\small decon}}$ (cf. Eq. (\ref{eq_11})) is calculated with a single computed random value $\Delta W_1$ while $\sigma_{\mbox{\small est}}$ chooses an optimal value after $I$ iterations. As observed in Fig.~\ref{comparison_sigma}, $\sigma_{\mbox{\small est}}$ with Dow Jones daily data from October 1928 to July 2011 and in all studied models is less noisy than $\sigma_{\mbox{\small decon}}$. The fluctuation values of the deconvoluted is three or four orders of magnitude larger than the fluctuation values of the three ML algorithms herein proposed. 

We also stress the fact that the SV model jointly with their parameters are chosen before starting the computation. The parameters can however be easily estimated beforehand using historical data~\cite{eisler2007volatility}. See for instance Refs. \cite{Shephard,barucci,comte,griffin,jongbloed,morimune,todorov,andresen,asai} for alternative procedures for reconstructing volatility being more or less dependent on the volatility model chosen. Some of these approaches also include the parameter estimation procedure within the volatility estimation. Others are mainly devoted to capture the long term memory of the volatility.

\section{Results and comparison between models}
\label{sec_simres}

We here study the probability density of the volatility, the conditional return, the Mean First Passage Time (MFPT) and the two most important correlations with time (volatility auto-correlation and return-volatility asymmetric correlation or leverage effect) along the three different SV models. Data to perform comparisons across the different models described in Sections~\ref{sec_behaviour}-\ref{sec_corr} corresponds to the Dow Jones daily data from October 1928 to July 2011 but Section~\ref{sec_diffindexes} extends the survey to other financial market indices.

The parameters we use for the numerical calculations are those given in literature to reproduce the DJI \cite{dragulescu,masoliver2006multiple,perelloijtaf} and they are summarized in Tab.~\ref{table_alphamkcoef}.

\begin{table}[t]
\caption{Parameters, measured in daily units, for the three SV Models.}
\label{table_alphamkcoef}
\vspace{0.3cm}
\begin{center}
\begin{tabular}{llll}
\hline\hline\noalign{\smallskip}
 & $k$ & $\alpha$ & $m$ \\
\noalign{\smallskip}\hline\noalign{\smallskip}
OU & $1.4\times10^{-3}$ & $5\times10^{-2}$ & $1.2\times10^{-2}$ \\
Heston & $2.45\times10^{-3}$ & $4.5\times10^{-2}$ & $8.62\times10^{-5}$ \\
ExpOU & $4.7\times10^{-2}$ & $1.82\times10^{-3}$ & $8\times10^{-3}$ \\
\noalign{\smallskip}\hline
\hline
\end{tabular}
\end{center}
\end{table}

\subsection{Behavior of our estimator}
 \label{sec_behaviour}

\begin{figure}
\vspace{-7cm}
\includegraphics[scale=0.45]{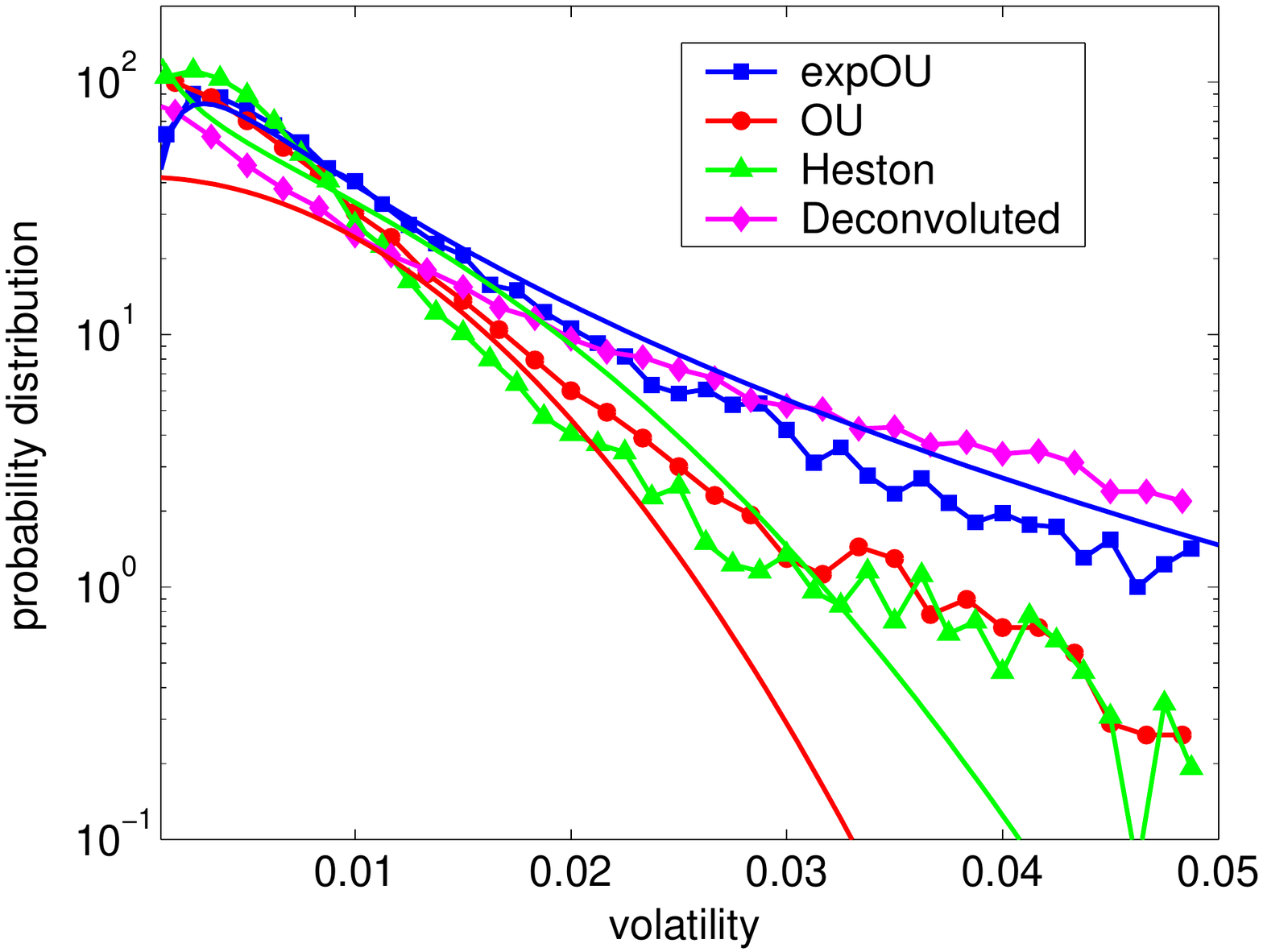}
\caption{Probability distribution of the different volatilities in semi-log scale: $\sigma_{\rm decon}$ (cf. Eq. (\ref{eq_11}) for the Dow Jones data and the $\sigma_{\rm est}$ for the expOU, the OU and the Heston cases (cf Eqs. (\ref{eq_6}), (\ref{estimation}) and Table \ref{table_fgh}). We also include theoretical stationary pdf forms for each model. The expOU seems to be the one that better corroborates theoretical pdf form.}
\label{pdf_volatility} 
\end{figure}

\begin{figure}
\vspace{-7cm}
\includegraphics[scale=0.45]{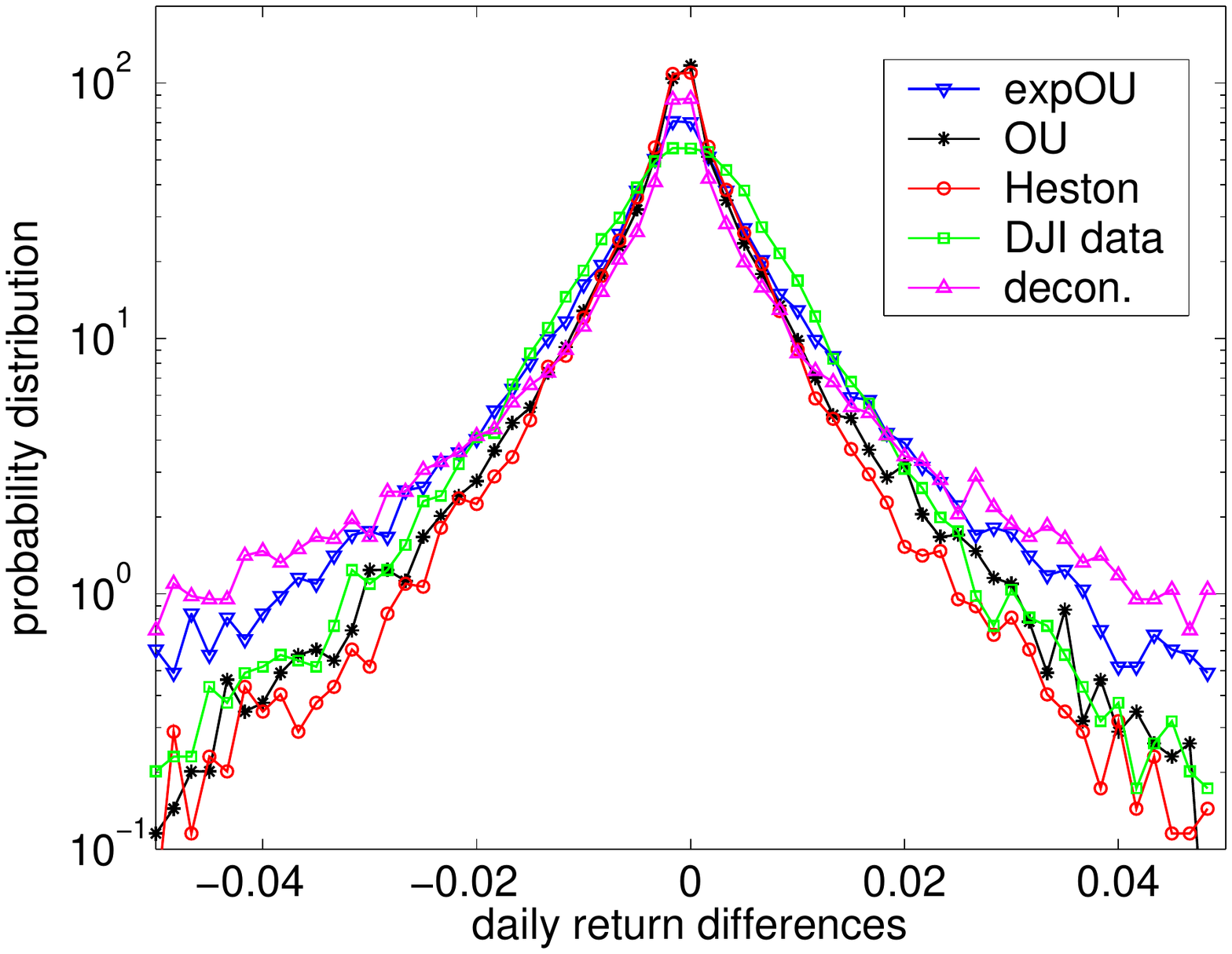}
\caption{Comparison between the probability density of the return differences $\Delta X$ calculated using Eq.~(\ref{eq_3}). We observe that the expOU model is the one that provides worse agreement with empirical data probably because of its high sensitivity of the parameter calibration.}
\label{pdfsreturn}
\end{figure}

In order to compare how our algorithm works on each model, we have first calculated the probability distribution of the different volatilities. Just for the sake of completeness we represent the stationary volatility probability density function (pdf) in Fig.~\ref{pdf_volatility} thus showing, as expected, that the form of the curves depends on the model choice. It should be noticed that we have used the absolute value of the volatility in the case of the OU model for the whole paper. Figure~\ref{pdf_volatility} also shows that best agreement between theoretical curve and empirical data points corresponds to the expOU case. Several studies in the literature have measured volatility stationary pdf~\cite{bouchaud2003theory,masoliver2006multiple,silva,stanley} and all of them suggest an exponential decay corresponding to a log-normal curve~\cite{masoliver2006multiple,stanley} or an inverse gamma distribution~\cite{bouchaud2003theory} at least with low frequency data. It shall however be noted that a very recent model with a two-dimensional diffusion process succeeds to provide an inverse gamma distribution~\cite{delpini} and it can be indeed interesting to apply the methodology to this new model.

We also compute artificial return fluctuations $\Delta X(t)$ for each model by multiplying $\sigma_{\mbox{\small est}}(t)$ with a Wiener noise realization as given by Eq.~(\ref{eq_3}). Doing that, we can somehow compare the daily zero-mean return pdf of the three SV models with the empirical data of daily returns $\Delta X(t)$. In Fig.~\ref{pdfsreturn}, we observe that the peak of empirical data $\Delta X(t)$ is not reproduced by any model. In Fig.~\ref{pdfsreturn} we see that the tails of the real $\Delta X(t)$ are similar to empirical data in all models. The differences among the models can be explained by the fact that the parameter estimation in each model has not been systematically optimized. We observe that the expOU model is the one that provides worse agreement with empirical data. A possible reason for that might be due to the fact that the expOU model has a multiplicative relation with the underlying random process $Y$ with $\sigma=f(Y)=m\exp(Y)$ and therefore needs a really accurate calibration (cf. Tab.~\ref{table_fgh}).

\subsection{Predictive power of the method}
\label{sec_predpower}

\begin{figure}[t]
\vspace{-7cm}
\includegraphics[scale=0.45]{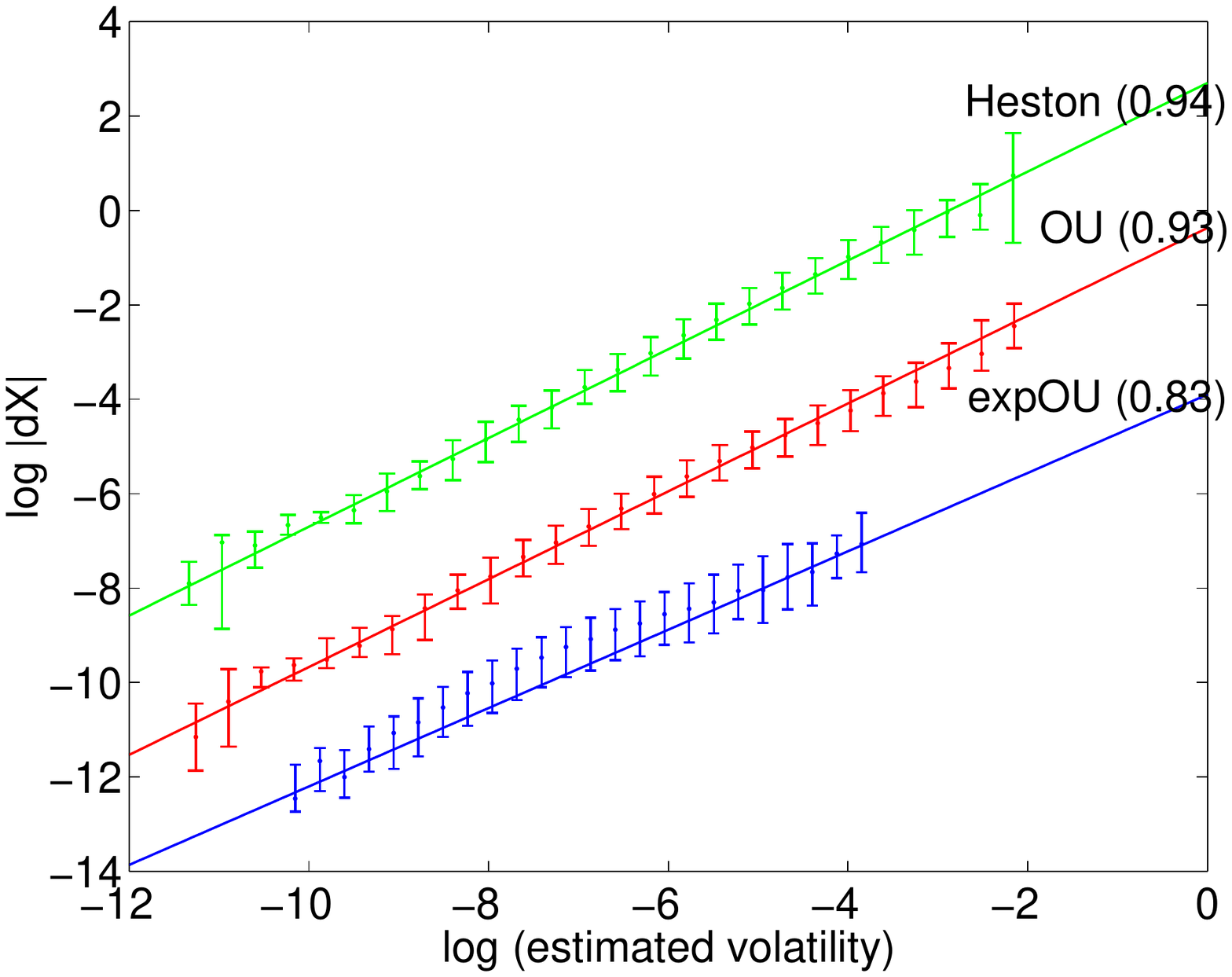}
\caption{Logarithm of the median of the empirical return differences as a function of the logarithm of the estimated volatility~(\ref{estimation}) for the three different models. All the models are shifted for better understanding. In brackets, we can find the value of the slope of the linear regression. The points represent the medians and the error bars are the first and third quartiles in the bins.}
\label{logdx_logsigma}
\end{figure}

This section aims to look for some inferred behavior in future absolute value zero-mean return based on the estimation of current value of volatility. We first consider the logarithm of Eq.~(\ref{eq_3})
\begin{equation}
 \label{eq_12}
\ln|\Delta X(t)| = \ln\sigma(t)+\ln |\Delta W_1(t)|,
\end{equation}
and we can now obtain the conditional median of the empirical $\ln|\Delta X(t)|=\ln|X(t+1\,\mbox{day})-X(t)|$ given we know $\ln\sigma(t)$ through our ML method. In such a case, we should have the following linear regression for the conditional median
\begin{equation}
 \label{eq_13}
M\left[\ln |\Delta X(t)|\Big|\ln\sigma(t)\right] = \ln\sigma(t)+ct,
\end{equation}
where $ct$ is a constant. In Fig.~\ref{logdx_logsigma} we plot this relationship using the three different models. We there however observe the slopes are not equal to 1. In this sense, Heston and OU model have the best performance although we should take into account that the performance might be very sensitive to the efficiency of the parameter estimation procedure.

\begin{figure}
\vspace{-7cm}
\includegraphics[scale=0.45]{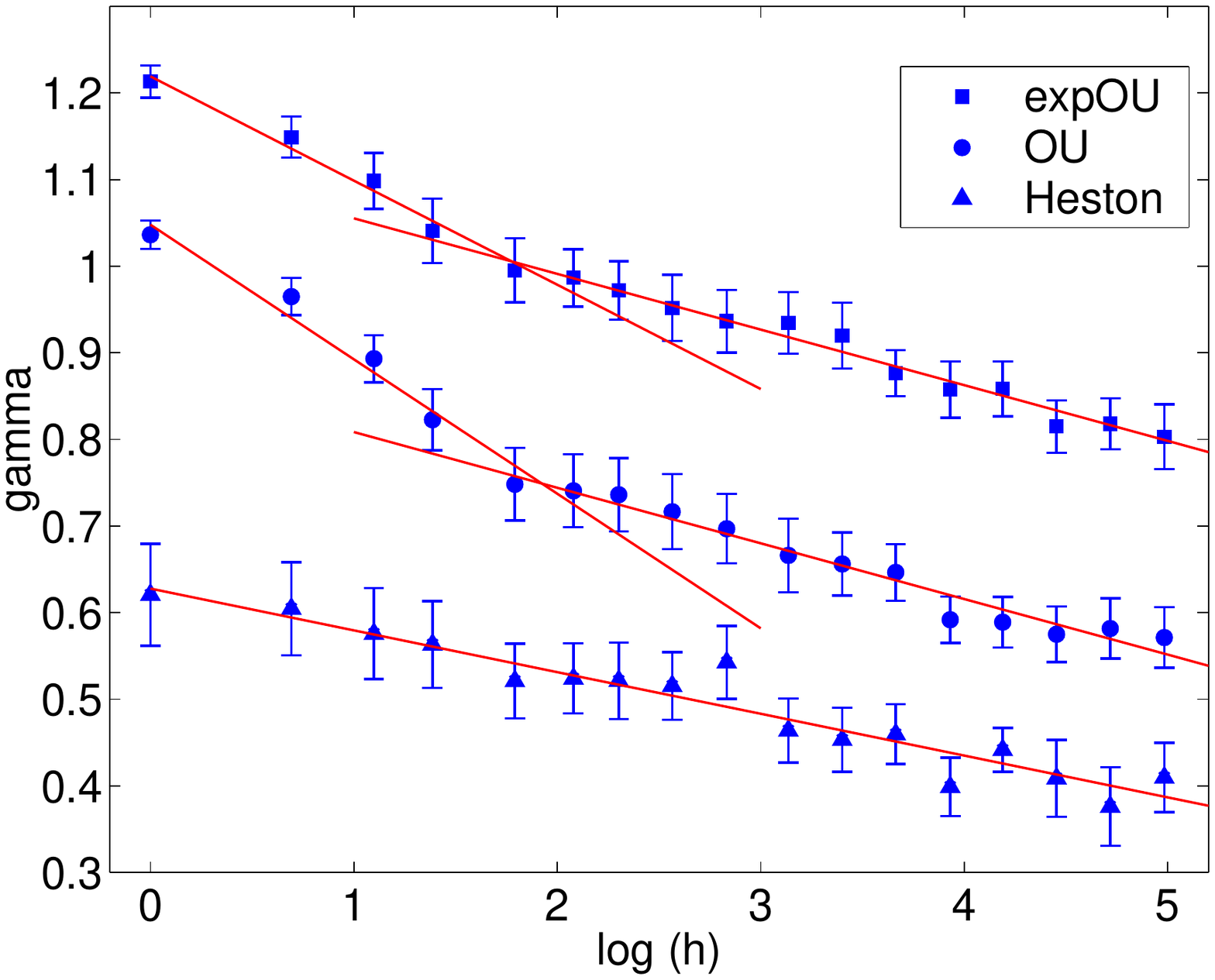}
\caption{Representation of the magnitude $\gamma(h)$ that appears in Eq. (\ref{eq_14}). The error bars correspond to the error on the slope of the regression of Fig. \ref{logdx_logsigma}. The data has been divided in two regimes in the case of the expOU and the OU models. All the plots are also shifted for sake of readability.}
\label{gamma_logh} 
\end{figure}

\begin{table}
\caption{Experimental values of the coefficients of Eq.~(\ref{eq_15}). The expOU and the OU models show a double time scale while the Heston model has a single time scale. Number 1 is valid for $h<7$ while number 2 applies for $h>7$.}
\vspace{0.3cm}
\label{table_abcoef} 
\begin{center}
\begin{tabular}{llllll}
\hline\hline\noalign{\smallskip}
 & expOU$^{\textrm{1}}$ & expOU$^{\textrm{2}}$ & OU$^{\textrm{1}}$ & OU$^{\textrm{2}}$ & Heston \\
\noalign{\smallskip}\hline\noalign{\smallskip}
$a$ & -0.12 & -0.064 & -0.15 & -0.064 &-0.048 \\
$b$ & 0.82 & 0.72 & 0.85 & 0.67 & 0.63 \\
\noalign{\smallskip}\hline\hline
\end{tabular}
\end{center}
\end{table}

In any case, we still have a linear regression measuring how big is going to be price fluctuation today based on yesterday's volatility level. One can go one step further and use the observed relationship between price fluctuations and volatility to forecast price changes amplitude at a longer time $t+h$ based on volatility at time $t$. A reasonable modification of the conditional median given by Eq.~(\ref{eq_13}) is
\begin{equation}
 \label{eq_14}
M\left[\ln |\Delta X(t+h)|\Big|\ln\sigma(t)\right]=\gamma(h)\ln\sigma(t)+ct,
\end{equation}
which was already proposed in Ref.~\cite{eisler2007volatility} but solely applied to the expOU case. We here therefore calculate $\gamma(h)$ in terms of time horizon $h$ for the expOU, the OU and the Heston cases. Figure~\ref{gamma_logh} shows a linear relation between $\gamma(h)$ and $\ln(h)$ for the three cases. We therefore propose the heuristic formula
\begin{equation}
 \label{eq_15}
M\left[\ln |\Delta X(t+h)|\Big|\ln\sigma(t)\right]=\left(a\ln h+b\right) \ln\sigma(t)+ct,
\end{equation}
where $a$ and $b$ are the coefficients of the regression. Table \ref{table_abcoef} shows the empirical values of the regression. Since we also observe a distinct behavior between short and long time horizon in the cases of the expOU and the OU models, we also provide two different regression parameters $a$ and $b$.

\subsection{Mean First-Passage Time}
 \label{sec_mfpt}

First-passage and extreme value studies have a long tradition of applications to physics, biology, chemistry, and engineering, all of them related to non equilibrium processes. This sort of events appear also to be important in the financial markets context as a valuable tool to calibrate risk in a more sophisticate manner than just providing the standard deviation. It also does represent an alternative and, in a way, improved method~\cite{masoliver} to the so-called Value at Risk~\cite{embrechts}. First-passage and other extreme value have already been analytically and empirically studied under the perspective of the here presented SV modeling~\cite{masoliver2007extreme,masoliver,mario}. In this section, we focus on the Mean First-Passage Time (MFPT) of the volatility which provides the average time spent by price fluctuations $|\Delta X|$ to cross a certain value $\lambda$. See Ref.~\cite{masoliver2007extreme} for a further theoretical input concerning the MFPT and the SV models herein studied.

We here want to extend the analysis with the use of our ML method instead of simply taking the absolute value of price returns as was done in Ref.~\cite{masoliver2007extreme}. In order to compare different models, we have to work with the dimensionless magnitude $L=\lambda/\sigma_s$ where $\sigma_s=\langle \sigma(t)\rangle_s$ is the mean of the estimated volatility in the stationary limit ($t\rightarrow \infty$). The expected stationary volatility \cite{masoliver2007extreme} for the expOU model is 
$$\sigma_s=m\exp{(k^2/4\alpha)},$$
for the OU model is $\sigma_s=m$, and for the Heston model is 
$$
\sigma_s=\frac{k\Gamma(2\alpha m^2/k^2+1/2)}{\sqrt{2\alpha}\Gamma(2\alpha m^2/k^2)}.
$$ 

\begin{figure}
\vspace{-7cm}
\includegraphics[scale=0.45]{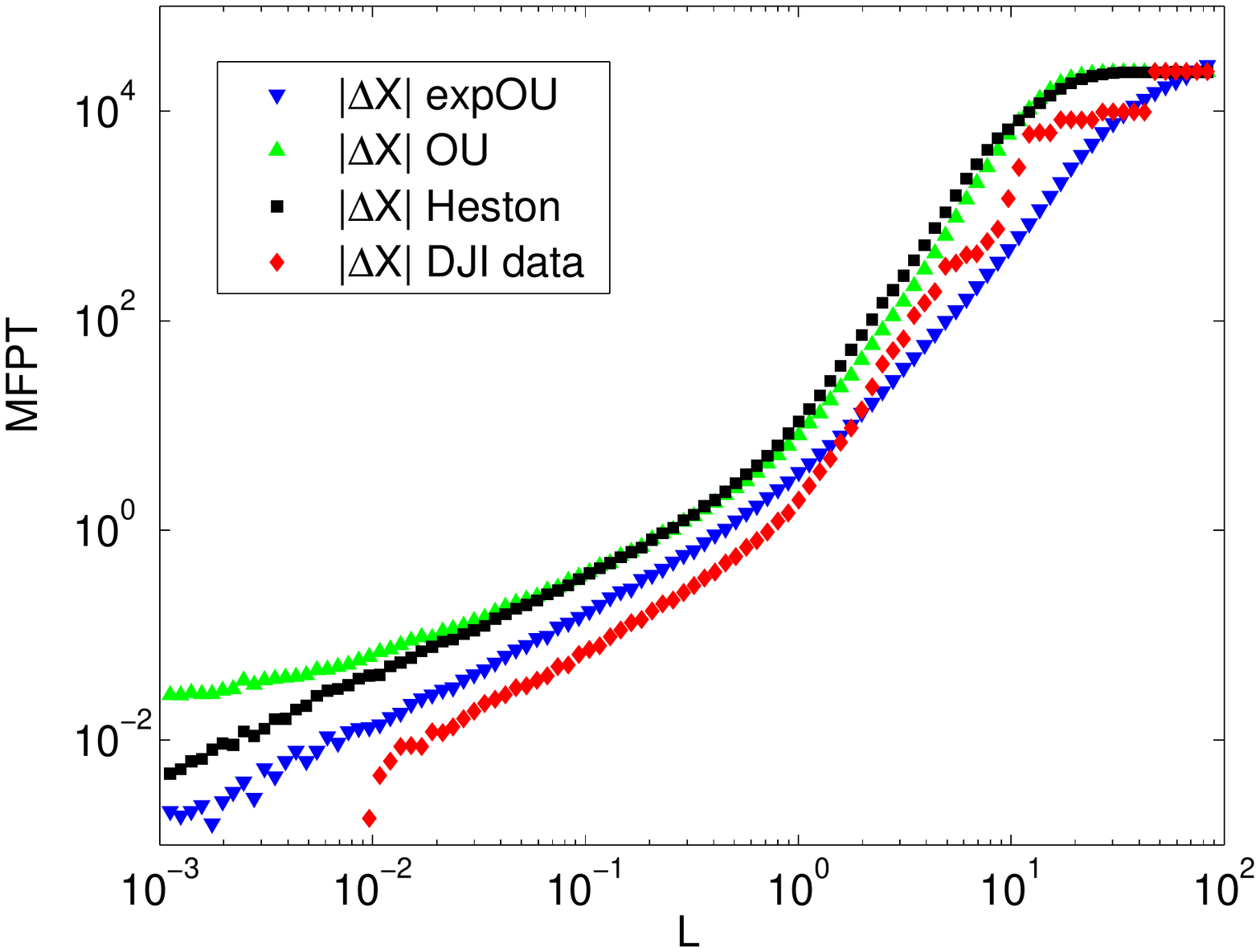}
\caption{MFPT of the return differences calculated using Eq.~(\ref{eq_3}). The estimated volatilities~(\ref{estimation}) of the expOU, the OU and the Heston models are compared with the Dow Jones $|\Delta X|$ data.}
\label{mfpt_dx} 
\end{figure}

The MFPT of the three models is computed with their own volatility estimation multiplied by an artificial Wiener random realization $\Delta W_1$. Figure \ref{mfpt_dx} compares the different results with a qualitative agreement with empirical data in all three cases. The expOU case appears to be the closest to the empirical MFPT curve. Figure~\ref{mfpt_dx} shows that the empirical MFPT results and the three artificial ones can all of them be roughly described by
\begin{equation}
\mbox{MFPT}(L)\simeq c\,L^{\beta}
\end{equation}
with exponent and coefficient that changes depending whether $L<1$ or $L>1$ as also shown in Ref.~\cite{masoliver2007extreme}. Their values are shown in Tab.~\ref{table_exponents_dX}.

\begin{table}
\caption{Scaling exponents $\beta$ of the MFPT of real data $\Delta X$ and artificial data of Heston, OU and expOU models. All the curves in Fig.~\ref{mfpt_dx} have a characteristic exponent for $L<1$ and another for $L>1$.}
\vspace{0.3cm}
\label{table_exponents_dX} 
\begin{center}
\begin{tabular}{lcccc}
\hline\hline\noalign{\smallskip}
 & expOU & OU & Heston & DJI data\\
\noalign{\smallskip}\hline\noalign{\smallskip}
$L<1$ & 1.1 & 0.8 & 1.0 & 1.3\\
$L>1$ & 2.4 & 3.1 & 2.9 & 2.9 \\
\noalign{\smallskip}\hline\hline
\end{tabular}
\end{center}
\end{table}

\subsection{Correlations}
 \label{sec_corr}

We now study how the ML approach keeps the main market time correlations that deeply and non-trivially involves volatility dynamics~\cite{bouchaud2003theory}. It is well-known that the volatility fluctuations have long memory correlation (over a year) and that volatility also shows negative and asymmetric cross-correlation with return changes (over several weeks), i.e. the leverage effect~\cite{bouchaud2003theory}. However, it is not clear whether the proposed method is able to provide these two different correlations.

\begin{figure}
\vspace{-7cm}
\includegraphics[scale=0.45]{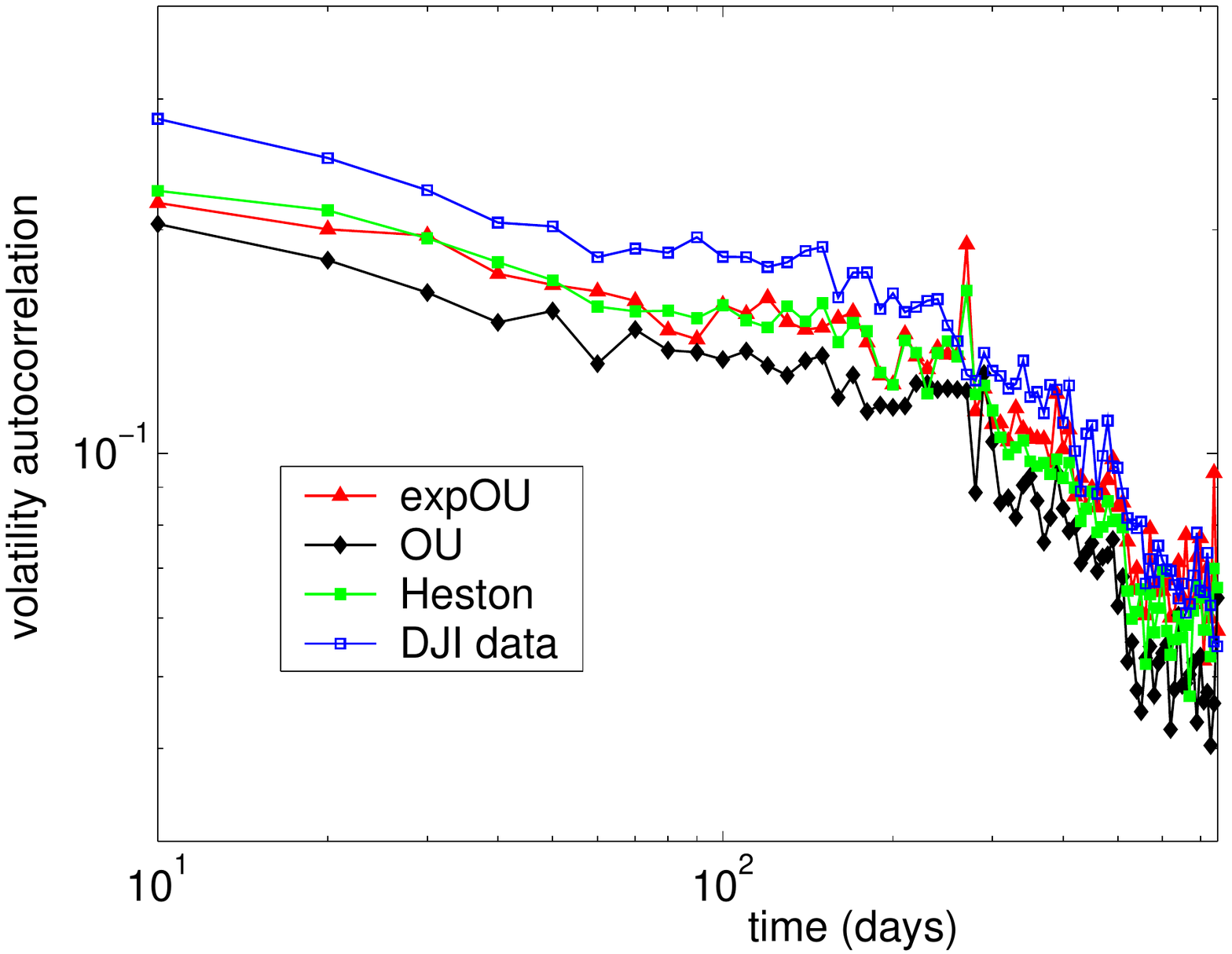}
\caption{Comparison between the autocorrelation of our volatilities jointly with the autocorrelation of the proportional $\sigma_{\mbox{\small prop}}$ provided by Eq.~(\ref{eq_10}).}
\label{volatility_autocorr}
\end{figure}

Figure~\ref{volatility_autocorr} shows how the volatility autocorrelation 
\begin{equation}
 \label{eq_15b}
\mathcal{C}(\tau) = \frac{\left\langle(\sigma(t+\tau)-\langle\sigma\rangle)(\sigma(t)-\langle\sigma\rangle)\right\rangle}{{\rm Var}[\sigma]}
\end{equation}
of each estimator is still significant for up to hundreds of days. It is important to stress that the OU and Heston models by themselves do not have this long range correlation since their mathematical expressions give an exponential decay for the volatility $\sigma$ in terms of a characteristic time scale $1/\alpha$ (see Ref.~\cite{perello2004comparison} and Tab.~\ref{table_fgh} for the meaning of this parameter). The expOU model is the only one that explains this long range effect with a cascade of exponentials~\cite{masoliver2006multiple}. Therefore, it can be said that the long-term memory is preserved due to the ML algorithmic method herein proposed. This feature manifests the robustness and effectiveness of the proposed method beyond the choice of the SV models been used.

\begin{figure}
\vspace{-7cm}
\includegraphics[scale=0.45]{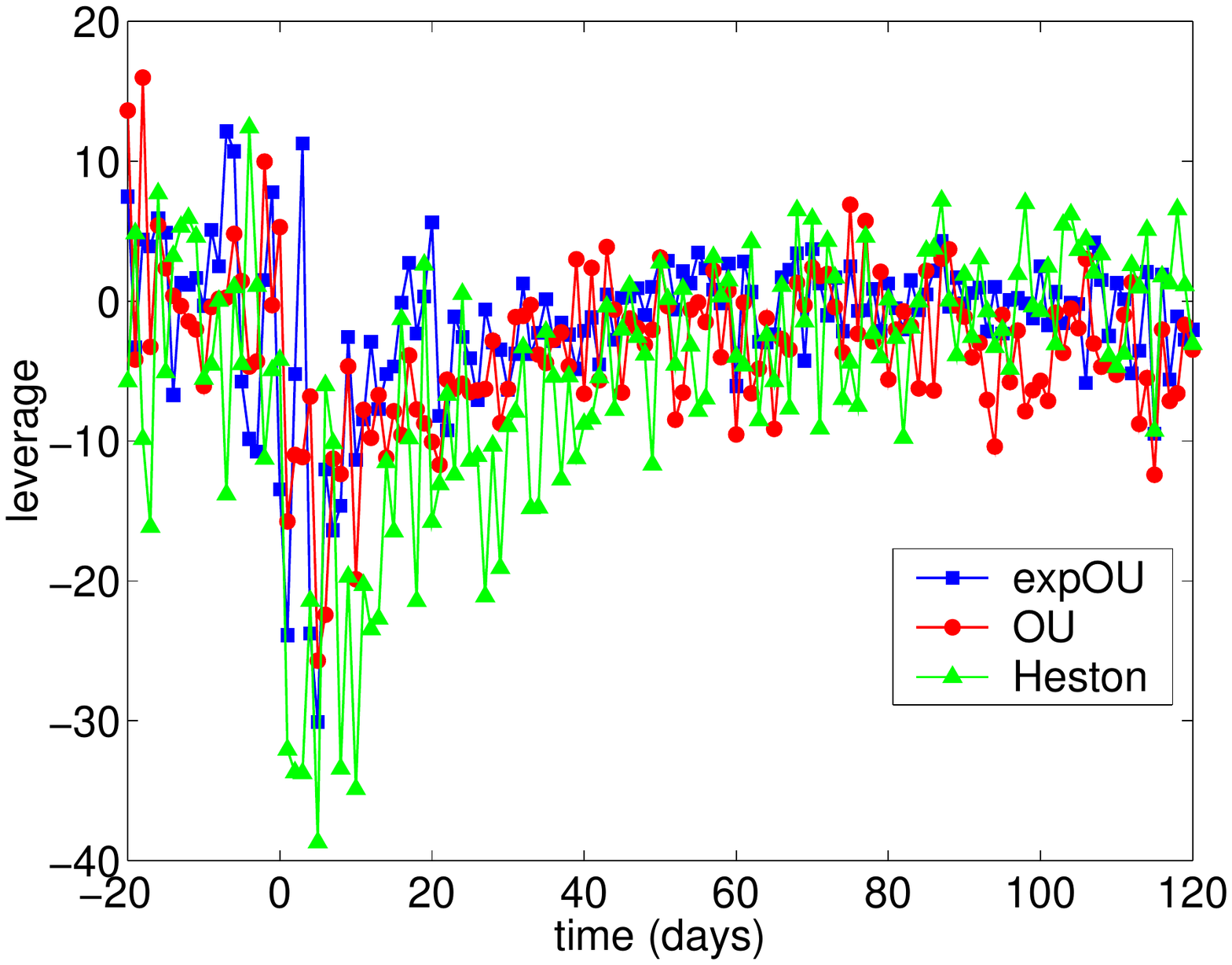}
\caption{Leverage correlations~(\ref{eq_16}) of the expOU, the OU and the Heston models. Volatilities are calculated using the ML method~(\ref{estimation}) and $|\Delta X|$ is articially computed combining Gaussian realizations of $\Delta W_1$ and taking the estimated volatility.}
\label{leverage_models} 
\end{figure}

We now focus on another important correlation with time. The so-called leverage effect~\cite{bouchaud2003theory} defined by
\begin{equation}
 \label{eq_16}
\mathcal{L}(\tau) = \frac{\left\langle \Delta X(t)\sigma(t+\tau)^2 \right\rangle}{\left\langle \sigma(t)^2 \right\rangle^2}
\end{equation}
measures the negative cross-correlation between price return fluctuations and volatility. Reference~\cite{leverage} shows that the three models are able to mathematically describe the empirical observation only if a non-zero and negative cross-correlation between $\Delta W_1$ and $\Delta W_2$ is considered (cf. Eq.~(\ref{eq_5})). Figure~\ref{leverage_models} shows the leverage correlation by first obtain the estimated volatility~(\ref{eq_16}) $\sigma_{\mbox{\small est}}$ and afterward compute the artificial return change $\Delta X$ by multiplying the estimated volatility by random realizations of $\Delta W_1$. We remind that the current ML algorithmic method has not considered correlation. However, the iterative procedure of the ML method is able to naturally provide the leverage effect in the three models as shown. It is important to stress the fact that we do not need to sophisticate our models by including the cross-correlation coefficient between $\Delta W_1$ and $\Delta W_2$ since the same ML procedure naturally includes the negative correlation between these random sources. Adding the effect of correlation between $\Delta W_1$ and $\Delta W_2$ represents adding more terms in Eq.~(\ref{eq_6}) and making the ML approach much less efficient in computational terms. The addition of this extra term would in any case provide redundant information to the maximization process. 

\begin{figure}
\vspace{-7cm}
\includegraphics[scale=0.45]{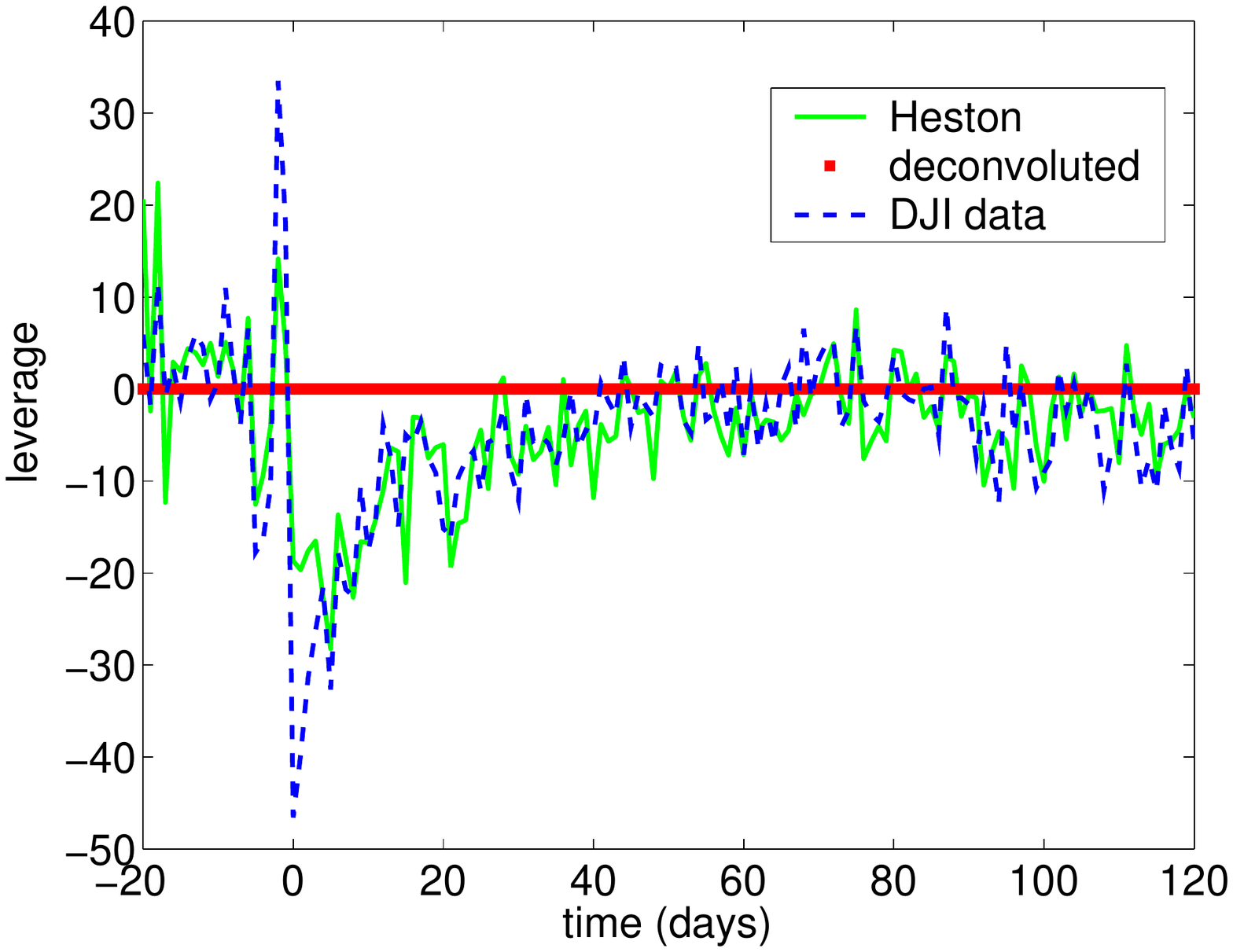}
\caption{Leverage correlations~(\ref{eq_16}) of the ML estimated volatility~(\ref{estimation}) for the Heston model compared with the deconvoluted procedure~(\ref{eq_11}) and the proportional volatility~(\ref{eq_10}).}
\label{leverage_heston} 
\end{figure}

Figure~\ref{leverage_heston} shows the leverage correlation of the Heston model as an illustrative example. It compares ML approach with other ways of extracting volatility from data. Figure~\ref{leverage_heston} demonstrates that ML approach gets same results as by using $\sigma_{\mbox{\small prop}}$ given by Eq.~(\ref{eq_10}) but it also shows how we lose the correlation if we take the deconvoluted $\sigma_{\mbox{\small decon}}$ given by Eq.~(\ref{eq_11}). Again, the result can be considered as a proof that our methodology is coherent and self-consistent. The other two models show very similar results as can be intuited in Fig.~\ref{leverage_models}.

\subsection{Different market indexes}
 \label{sec_diffindexes}

We have studied how our ML approach affects different SV models and we here would also like to verify if there is any difference between working with one stock market or another. Concretely, we have computed our estimation of the volatility for the following indexes: Dow Jones Industrial Average (DJI) (1928-2011), Standard and Poor's-500 (S\&P) (1950-2011), German index DAX (1990-2011), Japanese index NIKKEI (1984-2011), American index NASDAQ (1985-2011), British index FTSE-100 (1984-2011), Spanish index IBEX-35 (1993-2011) and French index CAC-40 (1990-2011). It is also important to stress that parameters used in each model are the ones from Dow Jones data and provided in Tab.~\ref{table_alphamkcoef} so in some sense there is now no over fitting due to the fact of extracting the parameter from the same data series we are analyzing. In all cases, the resulting $\Delta X$ time series satisfies the stylized facts that most of financial markets have in common~\cite{cont2001empirical,bouchaud2003theory}.

We first observe that all markets show an estimated volatility considerably less noisy than the deconvoluted one (cf. Eqs.~(\ref{estimation}) and~(\ref{eq_11})). The reduction of the oscillations can be quantified by the coefficient
\begin{equation}
 \label{eq_17}
\frac{\mbox{Var}(\sigma_{\mbox{\small est}})}{\mbox{Var}(\sigma_{\mbox{\small decon}})},
\end{equation}
whose order of magnitude depends on the stock data as shown in Tab.~\ref{table_variancescoef}.

\begin{table}[t]
\caption{Values of the coefficient $\mbox{Var}(\sigma_{\mbox{\small est}})/ \mbox{Var}(\sigma_{\mbox{\small decon}})$ for all the indexes. We show the values calculated using the expOU, the OU and the Heston models. }
\vspace{0.3cm}
\label{table_variancescoef}
\begin{center}
\begin{tabular}{lccc} 
\hline\hline\noalign{\smallskip}
 & expOU & OU & Heston \\
\noalign{\smallskip}\hline\noalign{\smallskip}
DJI & $8.6\times10^{-7}$ & $5.0\times10^{-7}$ & $2.5\times10^{-7}$ \\
S\&P & $3.0\times10^{-5}$ & $1.7\times10^{-5}$ & $6.3\times10^{-6}$ \\
DAX & $7.3\times10^{-7}$ & $3.4\times10^{-7}$ & $1.2\times10^{-7}$ \\
NIKKEI & $2.5\times10^{-6}$ & $1.8\times10^{-6}$ & $7.2\times10^{-7}$ \\
NASDAQ & $6.8\times10^{-6}$ & $4.9\times10^{-6}$ & $2.0\times10^{-6}$ \\
FTSE-100 & $3.7\times10^{-7}$ & $2.6\times10^{-7}$ & $8.4\times10^{-8}$ \\
IBEX-35 & $2.2\times10^{-4}$ & $1.9\times10^{-4}$ & $5.3\times10^{-4}$ \\
CAC-40 & $3.4\times10^{-6}$ & $2.1\times10^{-6}$ & $9.0\times10^{-7}$ \\
\noalign{\smallskip}\hline
\hline
\end{tabular}
\end{center}
\end{table}

\begin{figure}
\vspace{-7cm}
\includegraphics[scale=0.45]{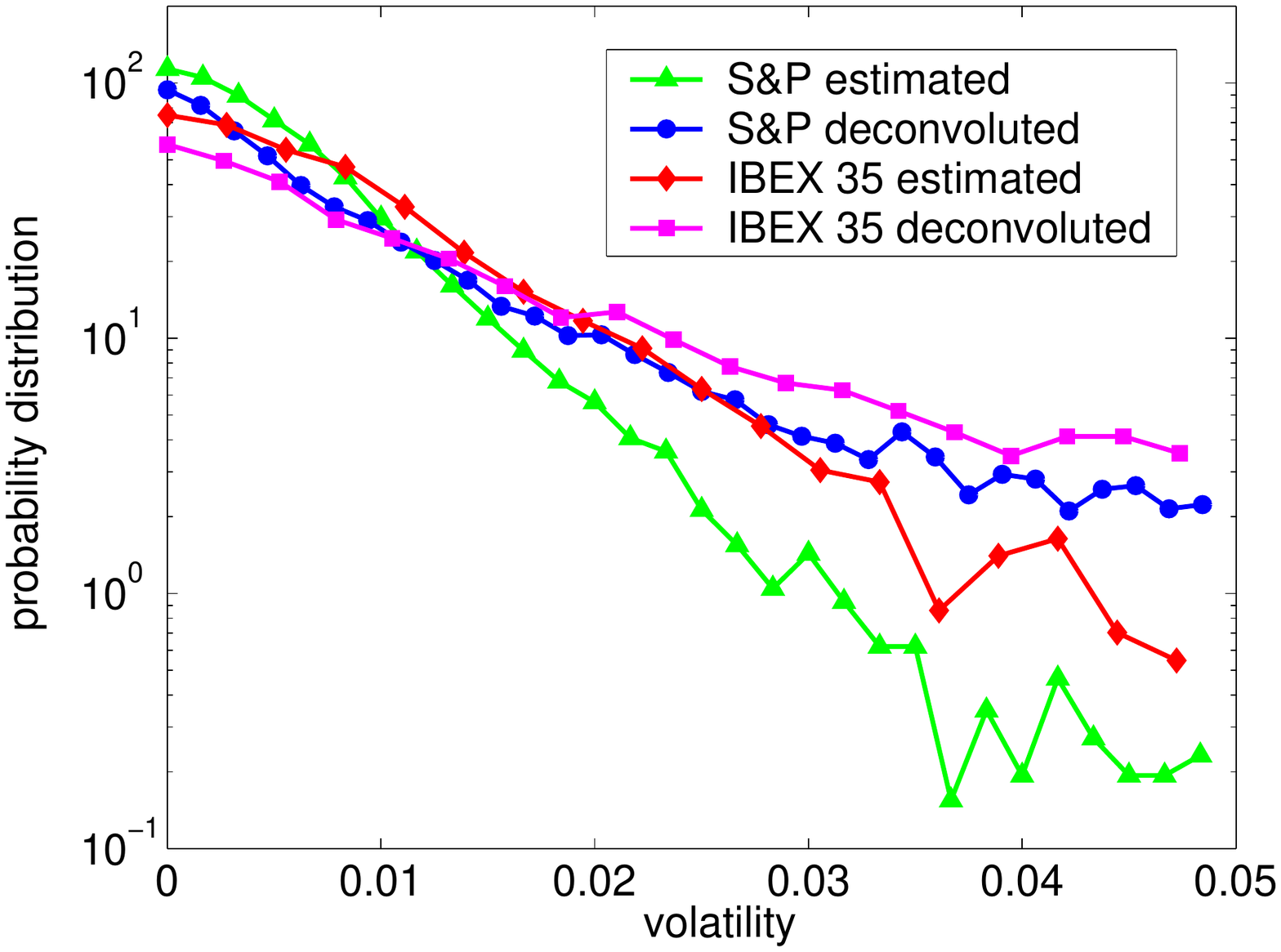}
\caption{Probability distribution of the estimated volatilities~(\ref{estimation}) of the Standard and Poor's-500 (S\&P) and the IBEX-35. We plot our estimated volatility for the Heston model jointly with the deconvoluted volatilities provided by Eq.~(\ref{eq_11}).}
\label{pdf_volatility_dif} 
\end{figure}

In Fig.~\ref{pdf_volatility_dif}, we plot the volatility pdf given by the Heston model for two different indexes. We notice the different width of the probability distribution of the two stocks because each market has a different volatility's range of values. We can again appreciate the reduction of the fluctuations achieved with our estimated volatility when compared with the deconvoluted volatility~(\ref{eq_11}). 

\begin{figure}
\vspace{-7cm}
\includegraphics[scale=0.45]{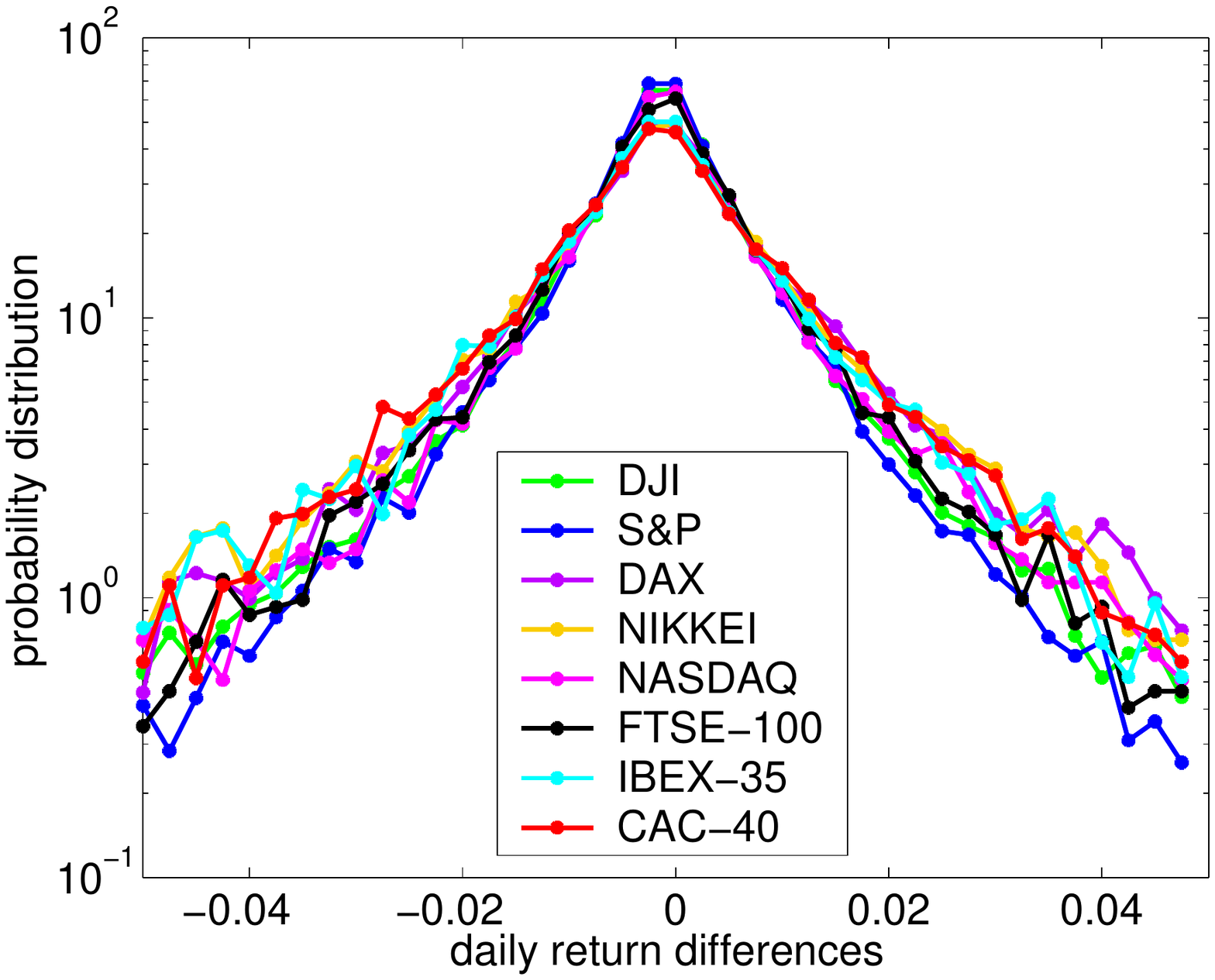}
\caption{Comparison between the probability density of the return differences $\Delta X$ artificially computed by using Eq.~(\ref{eq_3}) and by taking $\sigma_{\mbox{\small est}}$~(\ref{estimation}). The expOU model has been used in order to calculate the estimated volatility. All markets show similar aspect.}
\label{pdfsreturn_difst_expou} 
\end{figure}

Figure~\ref{pdfsreturn_difst_expou} shows the probability distribution of the artificially computed return differences with the estimated volatility of each stock. In this case, we have used the expOU model. As we expected, we see that the width of the curves depend on the stock market but behavior is qualitatively similar. This also similarly occurs to the Heston and OU models.

\begin{figure}
\vspace{-7cm}
\includegraphics[scale=0.45]{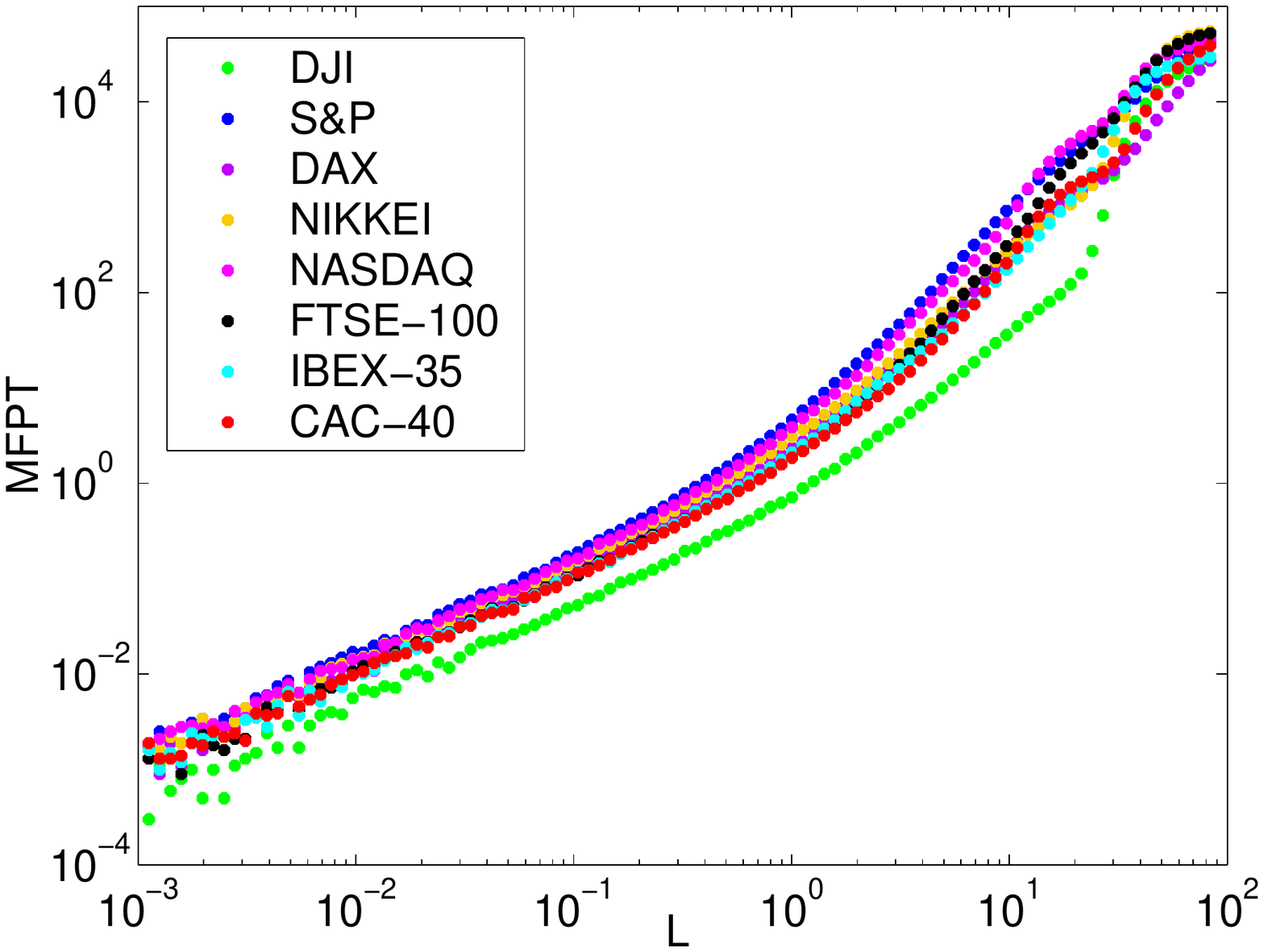}
\caption{MFPT of the absolute value of return differences calculated using artificial absolute value of return difference with the estimated volatility $\sigma_{\mbox{\small est}}$ (cf. Eqs.~(\ref{eq_3}) and~(\ref{estimation})). The estimated volatility has been computed using the expOU model.}
\label{mfptdX_difst_expou} 
\end{figure}

\begin{figure}
\vspace{-7cm}
\includegraphics[scale=0.45]{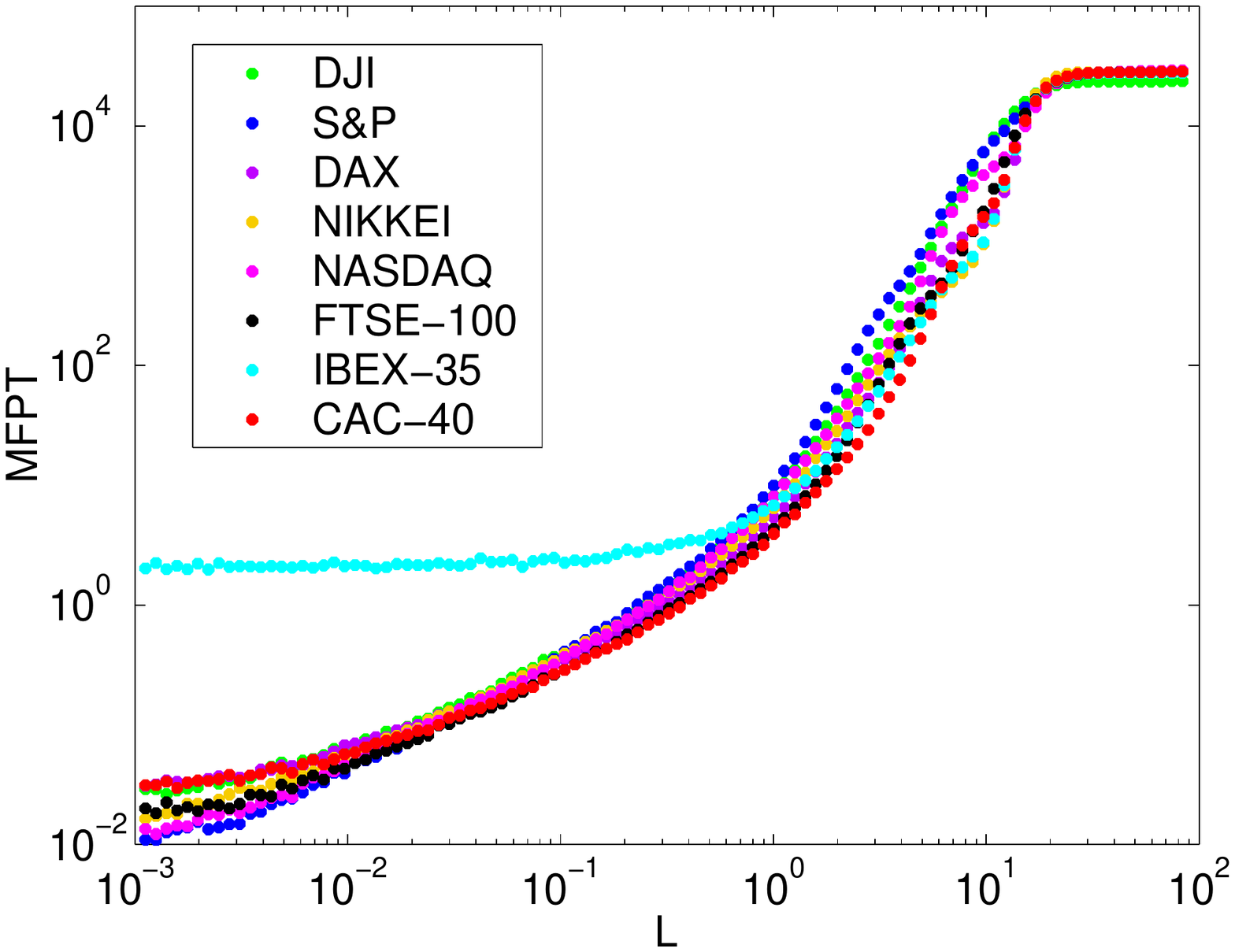}
\caption{MFPT of the absolute value of return differences calculated using artificial absolute value of return difference with the estimated volatility $\sigma_{\mbox{\small est}}$ (cf. Eqs.~(\ref{eq_3}) and~(\ref{estimation})). The estimated volatility has been computed using the OU model.}
\label{mfptdX_difst_ou} 
\end{figure}

\begin{figure}
\vspace{-7cm}
\includegraphics[scale=0.45]{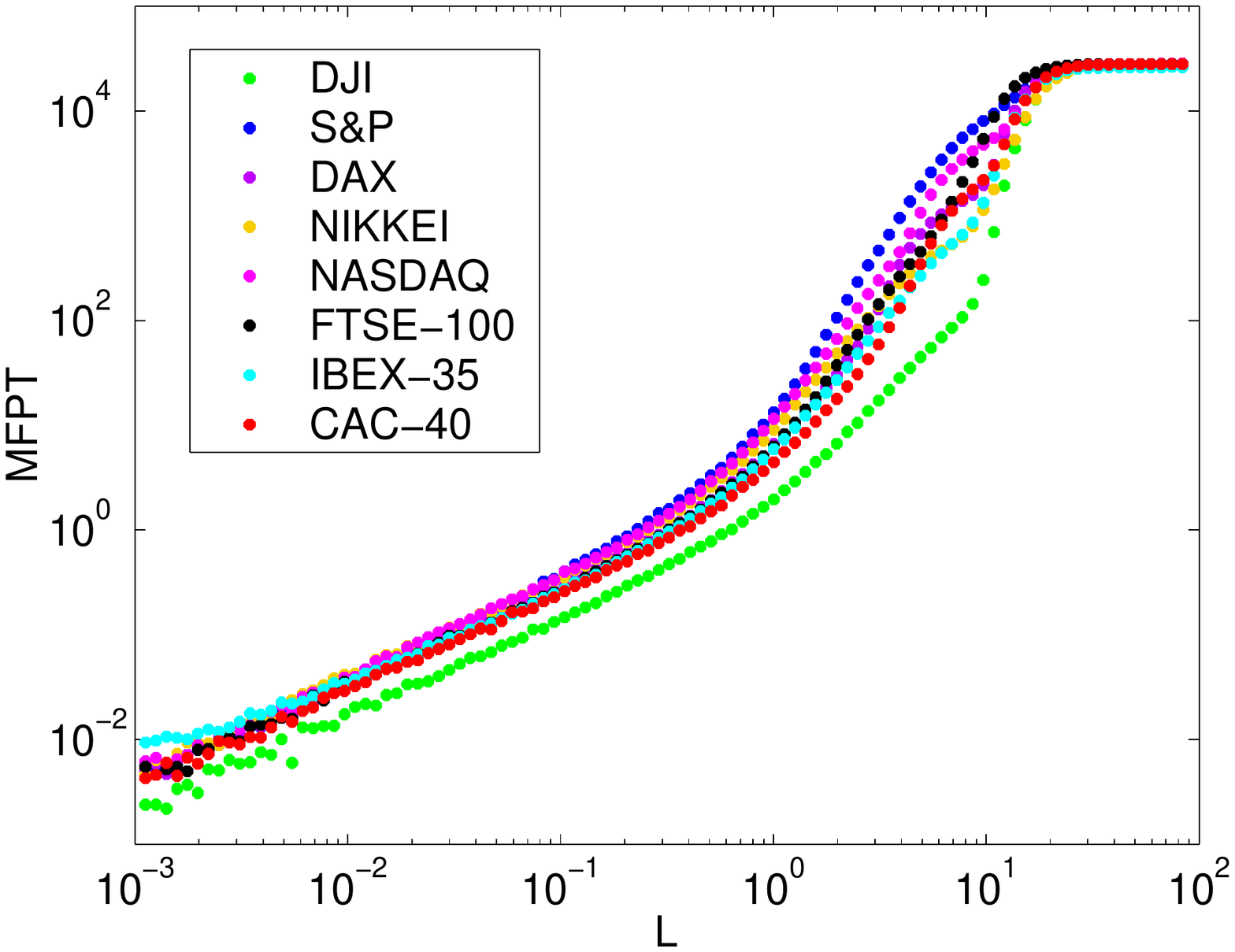}
\caption{MFPT of the absolute value of return differences calculated using artificial absolute value of return difference with the estimated volatility $\sigma_{\mbox{\small est}}$ (cf. Eqs.~(\ref{eq_3}) and~(\ref{estimation})). The estimated volatility has been computed using the Heston model.}
\label{mfptdX_difst_heston} 
\end{figure}

In order to study the extreme values of the indexes, we have calculated the MFPT for the absolute value of returns $|\Delta X|$ calculated using the estimated volatility. In top Fig.~\ref{mfptdX_difst_expou} we have plotted the evolution of this MFPT when the model used is the expOU. We observe the clear coincidence of all the stocks except the Dow Jones which has slightly smaller MFPT which is incidentally the market from where parameters are extracted. If we look at the the OU case as shown in Fig.~\ref{mfptdX_difst_ou} it is the IBEX stock index that shows a different behavior specially to the range of small threshold $L$. This can be justified by the fact that OU model allows for negative values of volatility while ML is just considering positive values of volatility. And the results for small $L$ will be the ones that can be more sensitive to this fact. Additionally the IBEX market is the one with smallest amount of data available. The Heston case shown in Fig.~\ref{mfptdX_difst_heston} recovers the nice collapse provided by the expOU model where the DJI again appears slightly shifted. In any case, and for the IBEX with the OU model single exception, a common pattern is observed.

\begin{figure}
\vspace{-7cm}
\includegraphics[scale=0.45]{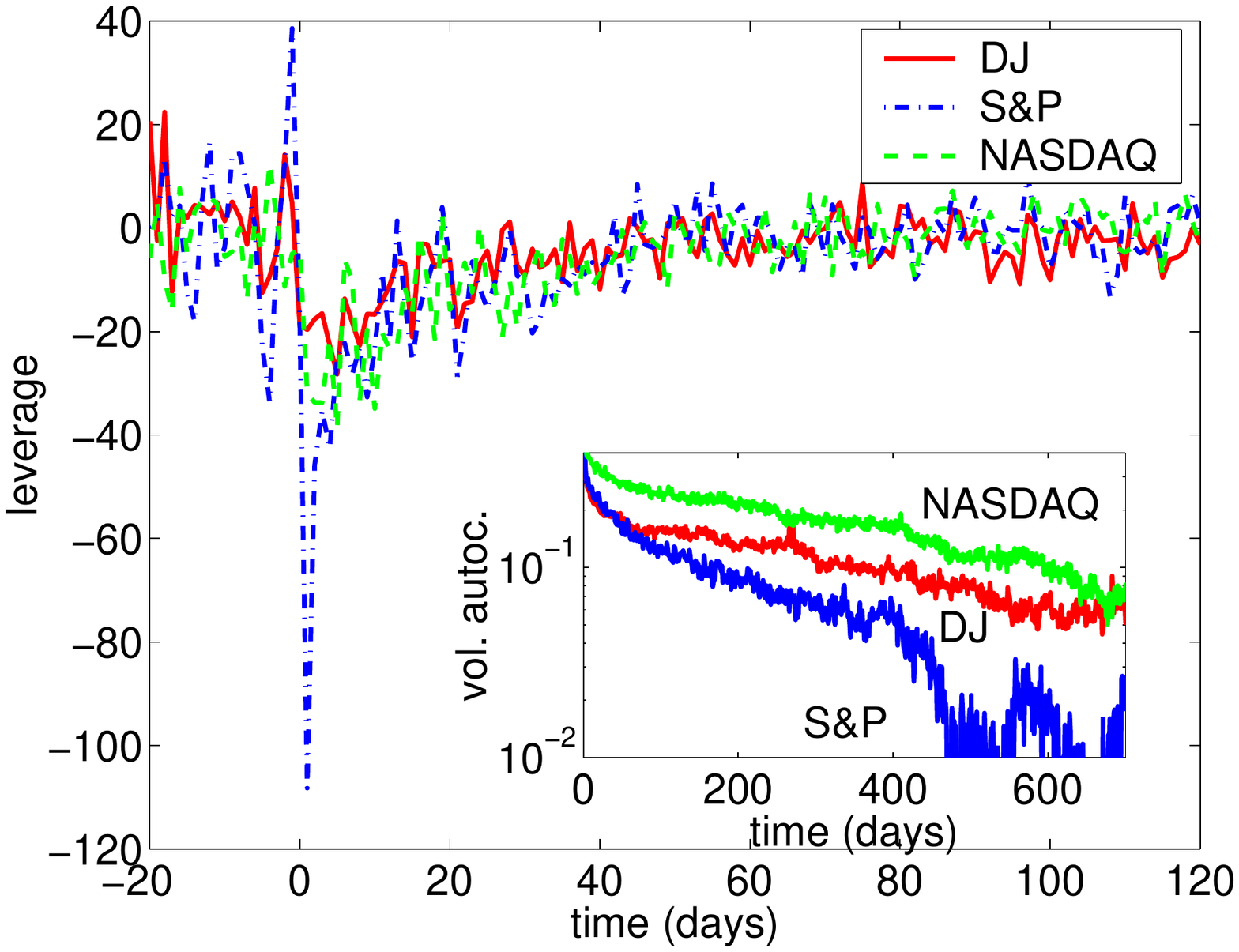}
\caption{Comparison between the leverage effect of the Dow Jones Industrial Average (DJ), Standard and Poor's-500 (S\&P), and NASDAQ. The inset Figure shows their volatility autocorrelation. The Heston model has been used in order to calculate the estimated volatility and the corresponding artificial return time series of each stock. The OU and expOU models and the rest of market indicies studied show identical results.}
\label{leverage_dif} 
\end{figure}

Finally, we show in Fig.~\ref{leverage_dif} that there are some stocks which manifest more leverage than others. As an example, the S\&P has bigger anti correlation than the Dow Jones. However, the important fact is that we find leverage in all markets. The same happens with the volatility autocorrelation because although the NASDAQ decays more slowly, all the stocks manifest significant autocorrelation for hundreds of days as expected~\cite{bouchaud2003theory}. Same results are found when we take the OU and expOU models instead of the Heston one.

\section{Conclusions}
 \label{sec_conclusions}
 
It is fairly known that the volatility is one of the main quantities in finance because it is a measure of price fluctuations and it gives information related to the risk of holding an asset. However, volatility is a magnitude which is not directly observable and one then needs to assume a given market model in order to infer the volatility value. Basic volatility estimation procedures have been presented and we have used a ML method that improves them since it is able to reduce noise and avoid bias in volatility signal.

We have applied the ML method by considering the most basic version of the expOU, the OU and the Heston SV uncorrelated models and we have compared them with the deconvoluted volatility showing big improvement in many aspects. We have observed that the fluctuations of the estimated volatility are smaller in all the models than in the deconvoluted estimation. The three models preserve the desired stationary volatility pdf for the volatility and keep the fat tail distribution for the price return changes. We have also found that all three models allow us to forecast future absolute value of returns with actual volatilities. We have also observed that the loss of forecast information has a double time scale in the expOU and the OU models.

Concerning the study of extreme events, we have found that our ML approach shows a nice concordance between the volatility MFPT estimated with the three SV models and the empirical MFPT. We have also focused on volatility's time correlations and we have observed that all the three models show the existence of significant volatility autocorrelation for hundreds of days although Heston and OU models does not include this property beforehand. The leverage correlation that crosses volatility and price return fluctuations is also nicely described by all three models even though the ML method is not considering correlation between returns and volatility fluctuations beforehand. All of these confirm the fact that methodology is robust enough without needing to improve the SV models or to provide more efficient ways of estimating the parameters of the model. However, ML approach with alternative models with same level of sophistication like the recent model by Delpini and Bormetti~\cite{delpini} deserves attention in future research.

Finally, we have applied same method to other stock indexes. Volatility's noise has been strongly reduced in all cases and we have corroborated that all the markets describe the several properties described before for the Dow Jones. The methodology therefore seems to be valid in a wide collection of financial market data.

\acknowledgements
Financial support from Direcci\'on General de Investigaci\'on under Contract No. FIS2009-09689 is acknowledged.


\end{document}